\documentstyle[11pt,epsfig]{article}
\begin{document}

\title{CHARGE RADII OF $\beta$-STABLE NUCLEI}

\author{G. K. NIE
\\
{\it Institute of Nuclear Physics, Ulugbek, Tashkent 702132, Uzbekistan}\\
galani@Uzsci.net}

\date{}
\maketitle

\begin{abstract}

In previous work it was shown that the radius of nucleus $R$ is determined by the
$\alpha$-cluster structure and can be estimated on the number of $\alpha$-clusters
disregarding to the number of excess neutrons. A hypothesis also was made that the
radius $R_m$ of a $\beta$-stable isotope, which is actually measured at electron
scattering experiments, is determined by the volume occupied by the matter of the
core plus the volume occupied by the peripheral $\alpha$-clusters. In this paper it
is shown that the condition $R_m = R$ restricts the number of excess neutrons
filling the core to provide the $\beta$-stability. The number of peripheral clusters
can vary from 1 to 5 and the value of $R$ for heavy nuclei almost do not change,
whereas the number of excess neutrons should change with the number of peripheral
clusters to get the value of $R_m$ close to $R$. It can explain the path of the
$\beta$-stability and its width. The radii $R_m$ of the stable isotopes with $12
\leq Z \leq 83$ and the alpha-decay isotopes with $84 \leq Z \leq 116$ that are
stable to $\beta$-decay have been calculated.

{{\it keywords}: nuclear structure; alpha-cluster model; charge radius; matter
radius; excess neutrons}

\end{abstract}

{\it{PACS} Nos.: 21.60.-n; 21.60.Gx; 21.60.Cs.}

\section{Introduction}

It is well known that the liquid drop model  gives a successful formula  to
calculate the charge radii of stable nuclei in dependence on the number of nucleons
in a nucleus $R_{ch} \sim A^{1/3}$. In Fig. 1 the experimental radii [1,2,3,4,5] of
stable isotopes are shown in dependence on $A$ and $Z$. The experimental errors
which in most of the cases are within an interval from 0.006 Fm to 0.060 Fm, are not
presented in the figure.

\begin{figure}[th]
\centerline{\psfig{file=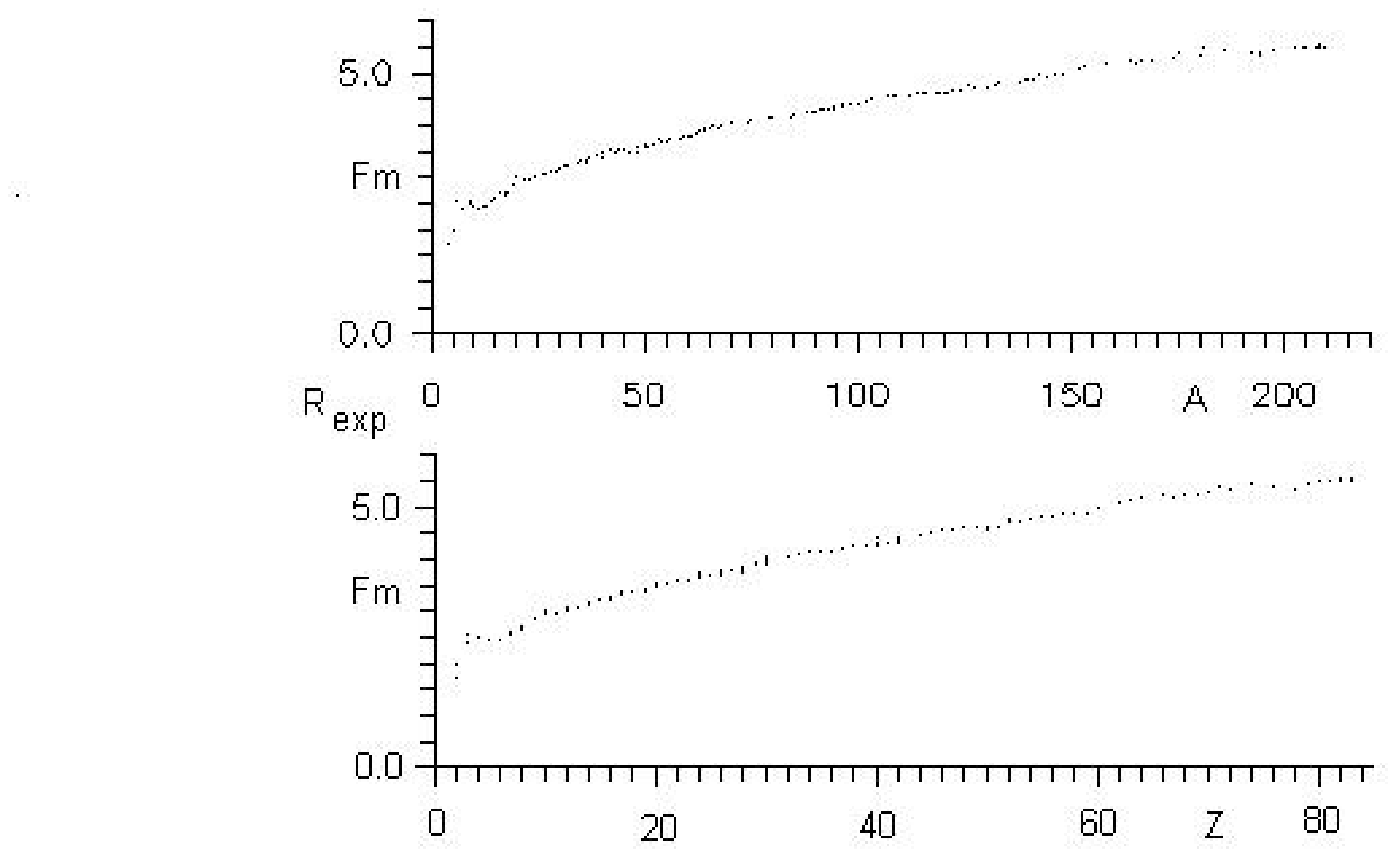,width=14cm}} \caption{Experimental radii of stable
isotopes in dependence on $A$ and  $Z$.}
\end{figure}

One can see that the values of the radii of nuclids of different isotopes of one
element in the graph of the dependance on $Z$ are gathered in short vertical
columns, whereas in the other graph the values are scattered around the line $R_{ch}
\sim A^{1/3}$.

The charge radii calculated by the following equation
\begin{equation}
R_{ch} = r_{ch} A^{1/3},\label{1}
\end{equation}
where  $r_{ch}$ denotes the average value of the charge radius of one nucleon in a
nucleus and\\
$ r_{ch} = 1.080$ Fm for the nuclei with   $6 \leq Z \leq 11 $,\\
$r_{ch}  = 1.012$ Fm for $12 \leq Z \leq 23 $, \\
$r_{ch}  = 0.955$ Fm for $24 \leq Z \leq 83 $ \\
have root mean square deviation from the experimental values of the most abundant
isotopes $<\Delta^2>^{1/2} = 0.067$ Fm.

The formula to calculate the radii of the nuclei in dependence on $Z$  or in
dependence on the number of $\alpha$-clusters $N_\alpha$ is

\begin{equation}
R_{ch} = R_{\alpha} N_\alpha^{1/3},\label{2}
\end{equation}
where $N_\alpha = Z/2$ in case of even $Z$ and in case of odd $Z$ the value of
$N_\alpha + 0.5$ is used. Using\\
$R_{\alpha} = R_{4\rm He} = 1.710$ Fm [1] for the nuclei with  $3 \leq N_\alpha \leq
5 $,\\
$R_{\alpha}  = 1.628$ Fm for $6 \leq N_\alpha \leq 11$,\\
$R_{\alpha}  = 1.600$ Fm for $12 \leq N_\alpha \leq 41$\\
gives $< \Delta^2>^{1/2} = 0.054$ Fm.

In the both cases three values of $r_{ch}$ and $R_{\alpha}$ have been used. The
changing the slope of the functions is explained by the formation of a core, which
starts growing from the nucleus with $Z =12$, $N_\alpha = 6$. For the nuclei with
$12 \leq Z \leq 22$, $6 \leq N_\alpha \leq 11$, the number of clusters of the core
is comparable with the number of peripheral clusters having the size of nucleus
$^4\rm He$. For the other nuclei with $Z \geq 24$, $N_\alpha \geq 12$, the number of
clusters of the core prevails and the mean radius of a core cluster is revealed as
to be $R_{\alpha} = 1.600$ Fm.

It was shown before that the size of a nucleus is determined by the $\alpha$-cluster
structure and the root mean square radius $R$ of a nucleus can be estimated by a few
ways on the number of $\alpha$-clusters in it [6,7], disregarding to the number of
excess neutrons. One of  the ways to estimate $R$ is (2).

To explain the paradox that both (1) and (2) can well describe the experimental
radii $R_{exp}$  the following hypothesis was proposed [8]. The $\beta$-stability,
which the most abundant isotopes belong to, is provided by the particular number of
excess neutrons in the isotope that is needed to fill in the space between the
volumes occupied by the charge and the matter of the alpha clusters of the core. The
size of an isotope $R_m$ is actually determined by the volume occupied by the matter
of the core $4/3\pi R^3_{m(core)}$ and the volume occupied by the charge of the
$N_{\alpha_{pr}}$ peripheral clusters

\begin{equation}
R_m^3 = R^3_{m(core)} + N_{\alpha_{pr}} R_{4\rm He}^3.\label{3}
\end{equation}
The $R_{m(core)}$ is calculated from

\begin{equation}
R^3_{m(core)} =  (N_\alpha - N_{\alpha_{pr}}) R_{m(\alpha)}^3 + \Delta N
r_{m(nn)}^3, \label{4}
\end{equation}
where $R_{m(\alpha)}$ stands for the matter radius of a core $\alpha$-cluster,
$\Delta N$ denotes the number of excess neutrons filling the core by pairs,
$nn$-pairs, and $r_{m(nn)}$ stands for the radius of one neutron of the pairs. The
condition

\begin{equation}
R_m = R\label{5}
\end{equation}
determines the particular  number of excess neutrons in the $\beta$-stable isotopes.
By fitting the values of the nuclear charge radii of the most abundant isotopes, the
value of the radius of one nucleon of $\alpha$-cluster matter of the core
$r_{m(\alpha)} =0.945$ Fm, which corresponds to $R_{m(\alpha)}= 1.500$ Fm, and the
value of $r_{m(nn)} = 0.840$ Fm were found [8].

In this paper the hypothesis that $R_m=R$ provides the $\beta$-stability is
developed. The root mean square charge radii $R$ of nuclei are calculated by a
phenomenological formula [6] with using charge radius of the core and the radius of
the peripheral $\alpha$-cluster position $R_p$ in the nucleus. The formula to
calculate $R_p$ was obtained from an independent analysis of the differences of
proton and neutron single particle binding energies in the framework of the
$\alpha$-cluster model based on $pn$-pair interactions.

In the article it is shown that for the nuclei with $Z \geq 24$ the value of $R$ is
almost independent of how many of the peripheral clusters $N_{\alpha_{pr}}$ are
placed on the surface of the core at the same radius $R_p$  from the center of mass
 $N_{\alpha_{pr}} = 1 \div 5$. However, in order to have the radius of an isotope
$R_m$ equal to its charge radius $R$ in the case of different numbers of peripheral
clusters the different number of excess neutrons is needed. This may explain why the
nuclids of different isotopes have close values of their radii and this also
explains the width of the narrow path of $\beta$-stability.

\section{Size of Nuclei}

The radius $R$ of a nucleus  is measured in electron scattering experiments as the
root mean square radius $<r^2>^{1/2}$ of the charge distribution. The $R_{ch}$ (2)
has meaning of $<r^3>^{1/3}$. The equality $<r^3>^{1/3} = <r^2>^{1/2}$ can be
provided by the only condition that the charge density distribution $\rho(r)=
const$, which is in an agreement with the results of analysis of the charge
distribution of heavy nuclei, see Fig. 6.1 [5]. The value of $\rho (r)$ is
approximately constant except the peripheral part with the approximately constant
thickness $t \approx 2.4\pm 0.3$ Fm for all nuclei from $^{16}\rm O$ to $^{208}\rm
Pb$.

To calculate the charge radius $R$ the following formula is used

\begin{equation}
N_\alpha<r^2> = (N_\alpha - N_{\alpha_{pr}} )<r^2>_{core} + N_{\alpha_{pr}}
<r^2>_p,\label{6}
\end{equation}
where $<r^2>_{core}$ denotes mean square radius of charge distribution of the core
and $<r^2>_p$ denotes the mean square radius of the charge distribution of
peripheral clusters. This equation is obtained from the definition of the root mean
square radius
\begin{eqnarray}
\nonumber N_\alpha <r^2> = 4\pi\int_{0}^{\infty}r^2 \rho(r) r^2dr,
\end{eqnarray}
where the spherically symmetrical charge density distribution $\rho(r)$,  normalized
as
\begin{eqnarray}
\nonumber 4\pi\int_{0}^{\infty}\rho(r) r^2dr = N_\alpha,
\end{eqnarray}
is the sum of the charge density distributions of the core $\rho_{core}(r)$ and the
peripheral clusters $\rho(r)_{\alpha_{pr}}$ with the corresponding normalization
\begin{eqnarray}
\nonumber 4\pi \int_{0}^{\infty}\rho(r)_{core} r^2dr = N_\alpha - N_{\alpha_{pr}},
\\
\nonumber 4\pi \int_{0}^{\infty}\rho(r)_{\alpha_{pr}} r^2dr =  N_{\alpha_{pr}}.
\end{eqnarray}

Eq. (6) is used for calculations of the charge and matter distributions of nuclei in
the wave function approach with the single-particle potential model assumed in the
shell model [9,10] in terms of the completed shells and the occupation numbers for
the nucleons of the last not completed shell.

Taking into account that for the nuclei with $Z \geq 24$, $N_\alpha \geq 12$, the
radius of one core $\alpha$-cluster $R_\alpha$ = 1.600 Fm  one can write for the
case when $N_{\alpha_{pr}} = 4$

\begin{equation}
< r^2>_{core}^{1/2} = 1.600 (N_{\alpha} - 4)^{1/3}.\label{7}
\end{equation}
The phenomenological formula [6] to calculate $R_p$ for the nuclei with $N_\alpha
\geq 12$, which has meaning of $<r^2>_{p}^{1/2}$, is
\begin{equation}
 < r^2>_{p}^{1/2}  = 2.168 (N_\alpha -4)^{1/3}.\label{8}
\end{equation}

Then the  equation (6) to calculate the charge radii $R$ becomes

\begin{equation}
N_\alpha R^2= (N_\alpha - 4) 1.600 ^2 (N_\alpha -4)^{2/3}  + 4 * 2.168^2 (N_\alpha
-4)^{2/3}\label{9}
\end{equation}
and for odd nuclei $R_1$, taking into account that $R_{p1}\approx R_p$ [6],
\begin{equation}
(N_\alpha +0.5)R^2_1 = (N_\alpha - 4) 1.600 ^2 (N_\alpha -4)^{2/3} + 4.5 *2.168^2
(N_\alpha -4)^{2/3}.\label{10}
\end{equation}

Eq. (9) and (10) were obtained [6]  in the framework of the alpha cluster model of
$pn$-pair interactions from analysis of  differences between the experimental values
of the binding energies of the last proton and neutron in the nuclei with $N=Z$. It
was shown that using (9) and (10) for the nuclei with $Z \geq 24$, $N_\alpha \geq
12$, gives a good fitting the experimental radii of the nuclids of the most abundant
isotopes with $<\Delta^2>^{1/2} = 0.050\rm Fm$, see, for example Fig. 2 [6]. The
deviation between $R_{ch}$ (2) and $R$ (9) and (10) for the nuclei with $24 \leq Z
\leq 116$  $<\Delta^2>^{1/2} = 0.028$ Fm.

To consider the cases with different amount of peripheral clusters, the equation (6)
should be rewritten as

\begin{equation}
N_\alpha R^2 = (N_\alpha - N_{\alpha_{pr}}) 1.600 ^2 (N_\alpha -
N_{\alpha_{pr}})^{2/3} + N_{\alpha_{pr}} 2.168^2 (N_\alpha -
N_{\alpha_{pr}})^{2/3}.\label{11}
\end{equation}

For the  nuclei with odd  $Z$

\begin{equation}
(N_\alpha +0.5) R^2_1 = N_\alpha R^2  +0.5 * 2.168^2 (N_\alpha -
N_{\alpha_{pr}})^{2/3}.\label{12}
\end{equation}

Let us analyze the function  $F(N_\alpha) = (R/(1.600^2N_\alpha^{1/3}))$. After
simplifying the expression one gets

\begin{equation}
F(N_\alpha) = (1 - N_{\alpha_{pr}}/N_\alpha)^{1/3}(1 - N_{\alpha_{pr}}/N_{\alpha} (1
- (2.168/1.600)^2)))^{1/2}.\label{13}
\end{equation}

In Fig. 2  the graph of the function is shown with four solid lines corresponding to
$N_{\alpha_{pr}} = 0, 3, 4, 5 $.  The small crosses indicate $F_{exp} = R_{exp}/
(1.600 N_\alpha^{1/3})$. The dashed lines denote  the  graphs of the function
$F(N_\alpha)$ with $N_{\alpha_{pr}} = 4$ and with replacement of the number 2.168
with the number 2.500 ( upper line)  and  1.500  (lower line), which shows that the
value 2.168 obtained from an independent analysis of binding energies of light
nuclei provide the function $F(N_\alpha)$ convergent to 1 at $N_\alpha  \geq 12$.

\begin{figure}[th]
\centerline{\psfig{file=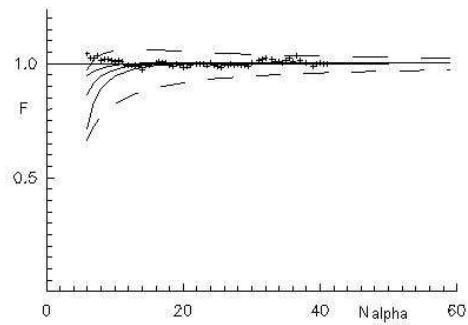,width=8cm}} \vspace*{2pt} \caption{Graphs of
function $F(N_\alpha)$ (13) with $N_{\alpha_{pr}} = 0, 3, 4 , 5$ (solid lines). The
graphs of $F(N_\alpha)$ with $N_{\alpha_{pr}} =  4$ (dashed lines) and  with the
values 2.500 (upper one)  and  1.500  (lower one) used instead of 2.168.}
\end{figure}

This means that $R$ for heavy nuclei does not significantly depend on the number of
peripheral clusters $N_{\alpha_{pr}}$. This explains why the radii calculated in the
$\alpha$-cluster model of completed shells $R_{shl}$ [6] are close to $R$. In the
model a nucleus consists of the $\alpha$-clusters of completed shells (the nuclei
$^{16}\rm O$, $^{20}\rm Ne$, $^{40}Ca$ and the nuclei placed in the right side of
the Periodic Table of Elements were taken as ones with completed shells) and the
number of clusters of the last not completed shell with the radius of their position
in the nucleus $R_p$ (8).

The difference between the equation to calculate $R_{shl}^2$ [6] and the equation to
calculate $R^2$ (9) is that the value $N_{\alpha_{pr}}$ varies from 0 to 5. For
example $N_\alpha R_{shl}^2 $ of nucleus $^{40}_{20} Ca$ is calculated as a sum of
$5 R_{^{20}_{10}\rm Ne}^2$ and $5 R_p^2$. The empirical values of $R_p$ in case of
the nuclei with $N_\alpha \leq 12$ were found in the framework of the model of
$pn$-pair interactions [6,7] on the values of differences between the
single-particle binding energies of last proton and neutron in the nuclei with
$N=Z$. The value of $R_{shl}$ for $^{52}_{24}\rm Cr$ is calculated with using the
radius of nucleus $^{40}_{20}\rm Ca$, as one with completed shells, and
$N_{\alpha_{pr}} = 2$.  For the nuclei with completed shells with $N_\alpha > 12$
the $R_{shl}$ is calculated by (2). The values of $R_{shl}$ have deviation with the
experimental data of the most abundant isotopes $<\Delta^2>^{1/2}=0.051$ Fm [6].

So one can say that a nucleus has a dense core with $N_\alpha - N_{\alpha_{pn}}$
clusters and some peripheral clusters placed at the equal distances from the center
of the core. The size R for the nuclei with $N_\alpha \geq 12$ does not depend
significantly of the number $N_{\alpha_{pn}} = 0 \div 5$.

The radius of a $\beta$-stable isotope $R_m$ is dependent on the volume occupied by
the core. It is implied that only those nuclids stay $\beta$-stable that have proper
amount of excess neutrons to make the radii $R_m$ equal to the size of the charge of
the nucleus $R$.

\section{Nuclear radii of Isotopes Stable to $\beta$-Decay }

The experimental radii $R_{exp}$ of different isotopes belonging to one element
differ from each other but the values are close and the differences are restricted
within several hundreds of Fm. The radius of a nuclid $R_m$ (3) is calculated from
the sum of cubic radii of its parts.

Let us consider first that there are four peripheral alpha clusters in the nuclei of
even $Z$ ( the mass number $A$ and the number of the excess neutrons $\Delta N$) and
four and half peripheral alpha clusters in case of nucleus with odd $Z_1 = Z + 1$
(the mass number $A_1$ and the number of excess neutrons $\Delta N_1$).

The excess neutron pairs are in the core and in case of odd $Z$ nuclei one excess
neutron is stuck with the single $pn$-pair. The suggestion is supported by the fact
that there is only one stable isotope $^{19}\rm F$ (the abundance of the isotope is
100\% [11]). The link between the single pair with spin $s=1$ and the excess neutron
with $s=1/2$ is explained by the spin correlation. The single neutron on the surface
of the nucleus is not 'seen' by the electrons scattered on the nucleus in the
experimental measurement of the radius $R_m$. Then the radius of isotopes $R_m$ and
$R_{m1}$ of nuclei $A$ and $A_1$ are calculated by the following equations [8]

\begin{equation}
R_m^3 = 4R_{4He}^3 + R_{m(\alpha)}^3 (N_\alpha - 4) +  r_{m(nn)}^3\Delta
N,\label{14}
\end{equation}
and
\begin{equation}
R_{m1}^3 = 4.5R_{4He}^3 + R_{m(\alpha)}^3 (N_\alpha - 4) + r_{m(nn)}^3(\Delta
N_1-1).\label{15}
\end{equation}

The equations (14) and (15) can be used to calculate charge radii $R_m$ of any
$\beta$-stable isotopes.

For the $\beta$-stable isotopes with $6 \leq Z \leq 11$ $R_m = R_{ch}$. For the
nuclei with $Z \geq 12$ $R_m$ is calculated by (14) and (15). For the nuclei with $6
\leq Z \leq 23$ the radii $R = R_{ch}$ (2) and for the nuclei with $Z \geq 24$ $R$
is calculated by (9) and (10).

The values of $R$ and $R_m$ for the stable nuclei with $Z \geq 12$ are presented in
Table 1.  The radii  are given in comparison with their experimental values. For the
unstable nuclei with $Z > 83$ the radii $R$ are given in Table 2 together with $R_m$
calculated for those $\alpha$-decay isotopes that are stable to the $\beta$-decay
[12].

\begin{center}{Table 1. Radii of nuclei. Abundance is
given in \%, radii in Fm}\end{center} {\begin{tabular}{cccccc|cccccc}\hline
$Z$&$A$&$Abn$ [11]&$R_{exp}$ [2,3,4]&$R$&$R_{m}$&$Z$&$A$&$Abn$ [11]
 &$R_{exp}$ [2,3,4]&$R$&$R_{m}$\\\hline
12&  24&   78.6& 2.985(30) & 2.958 &   2.991&  48& 112&  24.1 &             & 4.635& 4.595    \\
13&  27&  100.0& 3.06(9)   & 3.038 &   3.060&  49& 115&   95.8&    4.611(10)& 4.664& 4.634   \\
14&  28&   92.2& 3.14(4)   & 3.114 &   3.112&  50& 120&   33.0&    4.630(7) & 4.700& 4.684   \\
15&  31&  100.0& 3.24      & 3.186 &   3.195&  50& 118&  24.0 &             & 4.700& 4.666   \\
16&  32&   95.0& 3.240(11) & 3.256 &   3.224&  51& 121&   57.3&    4.63(9)  & 4.728& 4.703   \\
17&  35&   75.5& 3.335(18) & 3.322 &   3.302&  52& 130&   34.5&    4.721(6) & 4.762& 4.787   \\
18&  40&   99.6& 3.393(15) & 3.386 &   3.398&  52& 124&  4.61 &             & 4.762& 4.734   \\
18&  36&   0.34& 3.327(15) & 3.386&    3.328&  53& 127&  100.0&    4.737(7) & 4.790& 4.771   \\
19&  39&   93.2& 3.408(27) &3.448 &    3.402&  54& 132&  26.9&    4.790(22)& 4.800& 4.818   \\
20&  40&   97.0& 3.482(25) &3.507 &    3.427&  54& 130& 4.08 &             & 4.823& 4.801   \\
20&  42&  0.64 &           & 3.507&    3.460&  55& 133& 100.0&    4.806(11)& 4.851& 4.837   \\
21&  45&  100.0& 3.550(5)  & 3.565&    3.529&  56& 138&  71.7&    4.839(8) & 4.882& 4.883    \\
22&  48&   74.0& 3.59(4)   & 3.620&    3.583&   56& 136& 7.81 &             & 4.882& 4.866  \\
23&  51&   99.8& 3.58(4)   & 3.675&    3.647&  57& 139&  99.9&    4.861(8) & 4.910& 4.901    \\
24&  52&   83.8& 3.645(5)  & 3.618&    3.669&   58& 140&   88.5&    4.883(9) & 4.940& 4.913   \\
25&  55&  100.0& 3.680(11) & 3.649&    3.729&  58& 138&  0.25 &             & 4.940& 4.897    \\
26&  56&   91.7& 3.737(10) & 3.731&    3.750&  59& 141&  100.0&    4.881(9) & 4.967& 4.931   \\
27&  59&  100.0& 3.77(7)   & 3.763&    3.809&  60& 142&   27.1&    4.993(35)& 4.996& 4.943   \\
28&  58&   67.8& 3.760(10) & 3.836&    3.802&  61& 145&$\beta^+$&             & 5.023& 4.977   \\
28&  60&  26.2&  3.812(30) & 3.836&    3.829&  62& 152&   26.6&    5.095(30)& 5.052& 5.036   \\
29&  63&   69.1& 3.888(5)  & 3.868&    3.885&  62& 150& 7.47  &             & 5.052& 5.021    \\
30&  64&   48.9& 3.918(11) & 3.934&    3.904&  63& 153&   47.8&    5.150(22)& 5.078& 5.053   \\
30&  66&  27.8 & 3.977(20) &3.934 &    3.930&  64& 158&   24.9&    5.194(22)& 5.106& 5.095   \\
31&  69&   60.2&           & 3.966&    3.983&  64& 156&   20. &             & 5.106& 5.080         \\
32&  74&   36.7&           & 4.027&    4.050&   65& 159&  100.0&             & 5.131&5.112   \\
32&  72&  27.4 &  4.050(32)& 4.027&    4.026&   66& 164&   28.2&    5.222(30)& 5.158& 5.138  \\
33&  75&  100.0& 4.102(9)  & 4.058&    4.077&   66& 162&  25.5 &             & 5.158& 5.143  \\
34&  80&   49.8& 4.142(3)  & 4.115&    4.141&   67& 165&  100.0&    5.210(70)& 5.184& 5.170  \\
34&  76&  9.02 &           & 4.115&    4.094&   68& 166&   33.4&    5.243(30)& 5.210 &5.181  \\
35&  79&   49.5& 4.163(79) & 4.146&    4.143&   69& 169&  100.0&    5.226(4) & 5.235& 5.212  \\
36&  84&   56.9&           & 4.198&    4.205&   70& 174&   31.8&    5.312(60)& 5.260& 5.251  \\
36&  82&  11.56&           & 4.193&    4.189&   70 & 172&  21.8  &             & 5.260&5.237  \\
37&  85&   72.2&           & 4.230&    4.230&   71& 175&   97.4&    5.378(30)& 5.285& 5.267  \\
37&  87&  27.9 & 4.180     & 4.230&    4.252&   72& 180&   35.2&    5.339(22)& 5.310& 5.306  \\
38&  88&   82.6& 4.26(1)   & 4.278&    4.268&   72 & 178&  27.1 &             &5.310&5.292    \\
38&  86&  9.86&            & 4.278&    4.246&   73& 181&  100.0&   5.500(200)& 5.334& 5.321  \\
39&  89&  100.0& 4.27(2)    & 4.309&   4.292&   74& 186&    28.4&    5.42(7)  & 5.358& 5.345   \\
40&  90&   51.5& 4.28(2)    & 4.355&   4.308&   74 & 184&  30.6 &             & 5.358&5.349   \\
41&  93&  100.0& 4.317(8)   & 4.385&   4.352&   75& 187&    62.9&             & 5.383& 5.374   \\
42&  98&   23.8& 4.391(26)  & 4.429&   4.409&   76& 192&    41.0&    5.412(22)& 5.406& 5.412   \\
42&  96&  16.7 &             & 4.429&  4.388&   76 & 190&  26.4 &             & 5.406&5.399    \\
43&  99&$\beta^-$&           & 4.459&  4.431&   77& 193&    61.5&             & 5.430& 5.426   \\
44& 102&   31.6& 4.480(22)   & 4.500&  4.411&   78& 194&    32.9&    5.366(22)& 5.453& 5.436    \\
44& 100&  12.5 &             & 4.500&  4.446&  79& 197&   100.0&    5.434(2) & 5.477& 5.464    \\
45& 103&  100.0& 4.510(44)   & 4.529&  4.488&  80& 202&    29.8&    5.499(17)& 5.499& 5.500   \\
46& 108&   26.7& 4.541(33)   & 4.569&  4.521&  81& 205&    70.5&    5.484(6) & 5.522& 5.528   \\
46& 104&  11.0&              & 4.569&  4.500&  82& 208&    52.3&    5.521(29)& 5.544& 5.550    \\
47& 107&   51.4&  4.542(10)   & 4.598 &4.543&  82& 206&   25.1 &            & 5.544& 5.537   \\
48& 114&   28.9&    4.624(8) & 4.635&  4.613&  83& 209&   100.0&            &5.568&  5.564  \\
\hline
\end{tabular}}
\newpage

\begin{center} {Table 2. Radii of $\alpha$-decay isotopes stable to $\beta$-decay. $\Delta N$
stands for number of excess neutrons.}\end{center}
{\begin{tabular}{cccccc|cccccc}
\hline $Z$&$A$& $\Delta N$&$R_{ch}(2)$&$R(9,10)$&$R_{m}(14,15)$&$Z$&$A$& $\Delta
N$&$R_{ch} (2)$&$R(9,10) $&$R_{m}(14,15)$\\\hline
 84& 212&  44& 5.562&   5.589&   5.587&100& 256&  56& 5.895&   5.921&   5.929\\
 85& 215&  45& 5.584&   5.612&   5.613&101& 259&  57& 5.914&   5.942&   5.953\\
 86& 216&  44& 5.606&   5.633&   5.622&102& 260&  56& 5.934&   5.960&   5.961\\
 87& 219&  45& 5.627&   5.656&   5.649&103& 263&  57& 5.953&   5.981&   5.985\\
 88& 222&  46& 5.649&   5.676&   5.670&104& 266&  58& 5.972&   5.999&   6.004\\
 89& 225&  47& 5.670&   5.698&   5.696&105& 269&  59& 5.991&   6.019&   6.027\\
 90& 228&  48& 5.691&   5.718&   5.717&106& 272&  60& 6.010&   6.037&   6.046\\
 91& 231&  49& 5.712&   5.741&   5.742&107& 275&  61& 6.029&   6.057&   6.068\\
 92& 234&  50& 5.733&   5.760&   5.763&108& 278&  62& 6.048&   6.074&   6.087\\
 93& 237&  51& 5.754&   5.782&   5.788&109& 281&  63& 6.067&   6.094&   6.109\\
 94& 238&  50& 5.774&   5.801&   5.797&110& 284&  64& 6.085&   6.111&   6.128\\
 95& 241&  51& 5.795&   5.823&   5.821&111& 287&  65& 6.103&   6.131&   6.150\\
 94& 240&  52& 5.774&   5.801&   5.809&112& 290&  66& 6.122&   6.148&   6.168\\
 95& 243&  53& 5.795&   5.823&   5.833&113& 293&  67& 6.140&   6.167&   6.190\\
 96& 244&  52& 5.815&   5.842&   5.842&114& 296&  68& 6.158&   6.184&   6.208\\
 97& 247&  53& 5.835&   5.863&   5.866&115& 299&  69& 6.176&   6.203&   6.229\\
 98& 250&  54& 5.855&   5.882&   5.886&116& 302&  70& 6.194&   6.219&   6.247\\
 99& 253&  55& 5.875&   5.903&   5.910&117& 305&  71& 6.212&   6.239&   6.268\\\hline
\end{tabular}}
\vspace{1cm}

Eq. (14) and (15) mean that the nuclei with $Z$ and $Z_1$ have one core with the
same amount of excess neutrons in it $\Delta N_{core} = \Delta N = \Delta N_1-1$,
which leads to the relation between the mass numbers of neighboring nuclei $A_1 - A
= 3$. Indeed, there is always an isotope with even $Z = Z_1 -1$  with the mass
number $A$ less than $A_1$ on 3 in the $\beta$-stability path.

In case of stable nuclei $R$ and $R_m$ have been calculated for the most abundant
isotopes and for those nuclei that have mass numbers $A = A_1 -3$. For the unstable
nuclei the charge and matter radii have been calculated only for stable to
$\beta$-decay nuclei with $84 \leq Z \leq 116$ [12] with $A = A_1 - 3$. In the Table
the data for the nucleus with $Z=117$ is given also in a supposition that it is
$\beta$-stable.

In the Table the calculated values are given in comparison with the experimental
values. The deviation between $R_m$ and $R_{exp}$ is $<\Delta^2>^{1/2} = 0.048$ Fm.
One can see from the tables that the values of $R_m$ and $R$ are equal within a few
hundredths of Fm. The  mean deviation between values of $R$ and $R_m$ for the nuclei
with $ 11 \leq Z \leq 117$ $ <\Delta^2>^{1/2} = 0.031$ Fm.

Unlike the even nuclei the odd nuclei have only one, rarely two $\beta$-stable
isotopes [12]. The alpha cluster model based on $pn$-pair interactions [6,7]
provides some reasonable explanation for it. The analysis of the nuclear binding
energies made in the framework of the model shows that the single proton-neutron
pair on the nucleus periphery has six meson bonds with the six pairs of three nearby
$\alpha$-clusters with the energy about 15 MeV. This may constitute one big cluster
which consists of three and half or four and half $\alpha$-clusters. The four
clusters on the periphery may be preferable due to  $\alpha$-cluster's features to
have three meson bonds with the three nearby clusters, which corresponds to nucleus
$^{16}\rm O$, and the single $pn$-pair ties up the three nearby clusters with the
six meson bonds.

In Fig. 3  a distribution of  the number of excess neutrons of $\beta$-stable
isotopes on $Z$ is shown.

\begin{figure}[th]
\centerline{\psfig{file=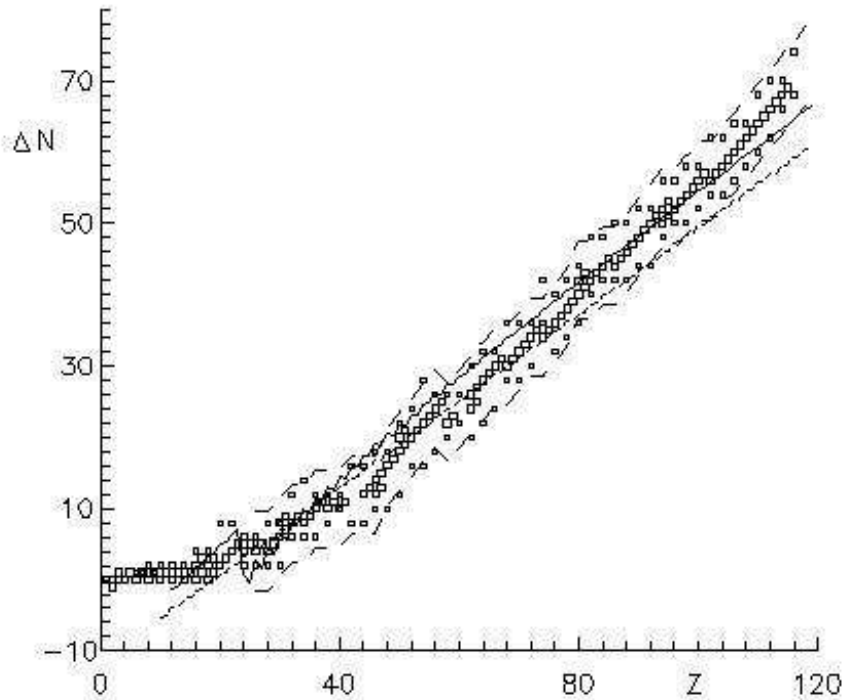,width=14cm}} \vspace*{2pt} \caption{The number of
excess neutrons of $\beta$-stable nuclei}
\end{figure}

The  chain of squares indicates the number of excess neutrons in the $\beta$-stable
isotopes with their  mass numbers $A = A_1 - 3$. The smaller squares scattered
around the chain indicate the numbers of the excess neutrons of even $\beta$-stable
nuclei with the biggest and the smallest $A$. The solid line indicates the number of
neutrons calculated by the following equation

\begin{equation}
\Delta N = (R^3 - R_{4\rm He}^3 N_{\alpha_{pr}} - R_{m(\alpha)}^3 (N_\alpha -
N_{\alpha_{pr}}))/r_{m(nn)}^3.\label{16}
\end{equation}

\begin{equation}
\Delta N_1 = \Delta N + 1 .\label{17}
\end{equation}

The short dashed line indicates the number of excess neutrons calculated by (16) and
(17) with $R = 1.600 N_\alpha^{1/3}\rm Fm$. The line $\Delta N$ corresponding to $R
= R_{shl}$, see Fig. 2 [8], is not presented here. In Fig. 3 it would be between the
solid and the dashed lines and for $Z > 85$ it goes above the chain of the squares
on several neutrons.

If in (16) one takes $R= R_m$ calculated by (14) and (15) for the isotopes $A$ and
$A_1$ being in the relation $A_1 = A + 3$, it is possible to estimate the width of
the $\beta$-stability path. In Fig. 3 two long dashed lines indicate $\Delta N$
calculated by (16) with $N_{\alpha_{pr}}= 6$ (lower one) and with the
$N_{\alpha_{pr}} = 2$ (higher line).

This fact that the stable isotopes of one nucleus have close values of radii is
illustrated on stable isotopes of $_{20}\rm Ca$. The size of the nucleus has been
estimated by a few ways. These are $R = R_{ch} = 3.507$ (2), Table 1, the value of
$1.600*10^{1/3} = 3.447$ Fm and $R_{shl} = 3.480$ Fm [6,7]. The value of $R_{shl}$
was calculated in the $\alpha$-cluster shell model, where nucleus $^{40}\rm Ca$ is
considered as a nucleus $^{20} \rm Ne$ plus five $\alpha$-clusters above it.

The radii $R_m$ of nuclei $^{40}\rm Ca$, $^{42}\rm Ca$, $^{44}\rm Ca$, $^{46}\rm
Ca$, $^{48}\rm {Ca}$  calculated (14) with the number of peripheral clusters 5, 5,
5, 3 and 2 the values of $R_m$ correspondingly are $R_{40\rm Ca} = 3.473$ Fm,
$R_{42\rm Ca} = 3.505$ Fm , $R_{44\rm Ca} = 3.537$ Fm, $R_{46\rm Ca}  = 3.481$ Fm,
$R_{48\rm Ca} = 3.468$ Fm. The radius  $R_{\rm 43Ca} = R_{\rm 42Ca}$, because the
last single excess neutron is not in the core consisted of zero spin objects, which
are $\alpha$-clusters and $nn$-pairs. It is supposed to be on the surface of the
nucleus and is not seen by the light charge particle scattered on the nucleus.

The experimental values of the radii of the isotopes are: for $^{40}\rm Ca$
(96.97\%) $R_{40\rm Ca} = 3.450(10)\div 3.482(25)$ Fm [2], 3.478 Fm [4], for
$^{42}\rm Ca$ (0.64\%) $R_{42\rm Ca} =3.508$ Fm [4], for ${^{44}\rm Ca}$ (2.06\%)
$R_{44\rm Ca} = 3.518$ Fm [4], for $^{46}\rm Ca$ (0.0033\%) $R_{46\rm Ca} = 3.498$
Fm [4], for $^{48}\rm Ca$ (0.185\%) $R_{48\rm Ca}=3.479$ Fm [4], 4.70 Fm and 3.51 Fm
[2], for $^{43}\rm Ca$ (0.145\%) $R_{43\rm Ca} = 3.495$ Fm [4]. Both experimental
and calculated values show that the radii of the stable isotopes are close within an
accuracy of the experimental data deviation of 0.03 Fm.

The data for the most explored nucleus $^{40}\rm Ca$ reveal the real accuracy of the
data obtained from the electron scattering measurement of the root mean square
radius of a nucleus. The deviation between the data taking into account the
measurement errors can be up to 0.06 Fm. It is because the results of the analysis
of the experimental data are still model dependent. It should define the minimal
deviation expected  to be obtained from theoretical calculations. From this point of
view the most appreciated measurement [4] has been carried out with one model of
analysis of data for several isotopes, revealing close but different values of the
radii.

\section{Conclusion}

The size of a nucleus $R$ can be estimated by a few ways in the framework of
$\alpha$-cluster model based on $pn$-pair interactions. These are $R_{shl}$ [6],
$R_{ch}$ (2), $R$ (9) and (10), and $R_m$ (14) and (15). All these estimations have
deviation with the experimental radii of the most abundant isotopes
$<\Delta^2>^{1/2} = 0.051$, 0.054, 0.050 and 0.048 Fm correspondingly. The average
deviation between the values is within 0.035 Fm

In calculations of root mean square radii $R_{shl}$ and $R$ a representation of a
core and a few peripheral clusters placed on the surface of the core at the same
distances $R_p$ from the center of mass is used. The number of peripheral clusters
$N_{\alpha_{pr}}$ is varied from 0 to 5 in calculations of $R_{shl}$ and
$N_{\alpha_{pr}} =4$ in case of $R$.

The values of $R_{shl}$, $R_{ch}$ and $R$ have been calculated on the number of
$\alpha$-clusters disregarding to the number of excess neutrons. The values of $R_m$
is calculated from a supposition that the core occupies some volume determined by
the alpha cluster matter and the matter of neutron-neutron excess pairs.

The $pn$-pair interaction model [6] provides an explanation of the fact that the
nuclei with odd $Z_1=Z+1$ have only one, rarely two $\beta$-stable isotopes $A_1$,
whereas the nuclei with even $Z$, have considerably bigger verity of $A$. The single
proton-neutron pair has six meson bonds with the three peripheral $\alpha$-clusters
with a large energy $\sim 15$ MeV, which constitutes one big peripheral cluster of
three and half or four and half $\alpha$-clusters with one excess neutron stuck with
the single $pn$-pair due to spin correlations between the pair with $s = 1$ and the
neutron with $s = 1/2$. Therefore the nuclei $A_1$ and $A = A_1-3$ have one core and
this chain determines the $\beta$-stability path.

It is shown that values of $<R$ do not depend significantly on the number of
peripheral clusters. For the case of different numbers of peripheral clusters the
different numbers of excess neutrons is needed to have the radius $R_m = R$. This
can determine the width of the $\beta$-stability path for the even nuclei.

The equations (14) and (15) can be used to calculate charge radii $R_m$ of any
$\beta$-stable isotopes. The radii $R_m$ of the most abundant isotopes are given in
comparison with their experimental values. The radii for the isotopes stable to
$\beta$-decay with $A$ and $A_1$ related with the equation $A = A_1 - 3$ for all
nuclei with $12\leq Z \leq 116$ are presented in Tables. The deviation between $R_m$
and $R$ $<\Delta^2>^{1/2} = 0.031 Fm$.

For calculation of the radii two parameters have been used. This is the matter
radius of a core $\alpha$-cluster 1.500 Fm and the radius of one neutron of the core
$nn$-pairs 0.840 Fm. The values may differ for various nuclei. However, as it is
shown here, they do not change considerably.

\end{document}